\RequirePackage[mathlines]{lineno} 
\documentclass[aps,prd,twocolumn,showpacs,amsmath,amssymb]{revtex4-1}
\usepackage[utf8]{inputenc}

\usepackage[hidelinks]{hyperref}    

\usepackage{epsfig}
\usepackage{graphicx}
\usepackage{dcolumn}
\usepackage{bm}
\usepackage{overpic}
\usepackage{subfigure}
\usepackage{float}
\usepackage{color}
\usepackage{tikz}
\usepackage{amsmath}
\usepackage{mathcomp}
\usepackage{units}

\usepackage{tikz}
\usetikzlibrary{positioning,arrows}
\usetikzlibrary{decorations.pathmorphing}
\usetikzlibrary{decorations.markings}
\tikzset{
fermion/.style={thick,draw=red, postaction={decorate},
decoration={markings,mark=at position .5 with {\arrow[red]>{triangle 45}}}},
anti-fermion/.style={thick,draw=blue, postaction={decorate},
decoration={markings,mark=at position .5 with {\arrow[blue]<{triangle 45}}}},
photon/.style={decorate, decoration={snake}, draw=black},
hadron/.style={thick,draw=black, postaction={decorate},
decoration={markings,mark=at position .5 with {\arrow[black]>{triangle 45}}}},
gluon/.style={decorate, draw=magenta,
decoration={coil,segment length=4}}
}

\newcommand{\mevcc}{\,\unit{MeV}/c^2}

\newcommand{\Dsp}{\ensuremath{D^+_s}}
\newcommand{\Dsm}{\ensuremath{D^-_s}}

\newcommand{\DsSTp}{\ensuremath{D^{*+}_s}}
\newcommand{\DsSTm}{\ensuremath{D^{*-}_s}}

\newcommand{\BESIIIorcid}[1]{\href{https://orcid.org/#1}{\hspace*{0.1em}\raisebox{-0.45ex}{\includegraphics[width=1em]{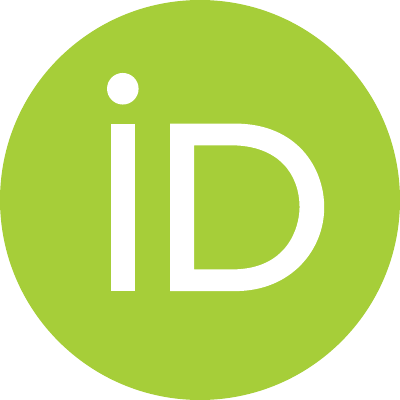}}}}

\let\oldequation\equation
\let\oldendequation\endequation
\renewenvironment{equation}
{\linenomathNonumbers\oldequation}
{\oldendequation\endlinenomath}

\begin{document}
\parskip=5pt plus 1pt minus 1pt

\title{ \boldmath First evidence for $D_s^+ \to f_1(1420) e^+\nu_e$ and search for $D_s^+ \to f_1(1285) e^+\nu_e$ }
\date{\today}

\begin{abstract}
Using $e^+e^-$ collision data corresponding to an integrated luminosity of 7.33~${\rm fb^{-1}}$ recorded by the BESIII detector at center-of-mass energies between 4.128 and 4.226~${\rm GeV}$,
we present the first search for the semileptonic decays $D^+_s\to\ f_1(1420)e^+\nu_e$ and $D^+_s\to\ f_1(1285)e^+\nu_e$.
The first evidence for the decay $D^+_s\to\ f_1(1420)e^+\nu_e$ is found with a statistical significance of 3.4$\sigma$, and its product branching fraction $\mathcal{B}(D^+_s\to\ f_1(1420)e^+\nu_e)\cdot\mathcal{B}(f_1(1420)\to\ K^+K^-\pi^0)$ is determined to be $\rm (4.5^{+2.0}_{-1.7}(stat) \pm0.4(syst)) \times 10^{-4} $, corresponding to an upper limit of $7.6 \times10^{-4}$ at the 90\% confidence level. No significant signal of the decay $D^+_s\to\ f_1(1285)e^+\nu_e$ is observed and the upper limit on the product branching fraction is set to be $\mathcal{B}(D^+_s\to\ f_1(1285)e^+\nu_e)\cdot\mathcal{B}(f_1(1285)\to\ \pi^+\pi^-\eta) < 1.7\times10^{-4}$ at the 90\% confidence level.

\end{abstract}

\maketitle
\section{Introduction}
\label{sec:introduction}

Within the Standard Model (SM), semileptonic (SL) decays of charmed mesons serve as an excellent laboratory for investigating the interplay between weak and strong interaction dynamics in the charm sector~\cite{LHBLyu2021}. Previously, SL decays of charmed mesons into pseudoscalar or vector mesons have been extensively studied. However, SL decays of charmed mesons involving axial-vector mesons in the final state remain relatively underexplored, both experimentally~\cite{PDG2024,KeBaiQ2023} and theoretically~\cite{Cheng2017pcq,397991989,5227831995,790360042009,S0217751X16501165}.

Understanding these decays requires precise knowledge of the structure and dynamics of the axial-vector mesons themselves. However, the enigmatic nature of axial-vector particles, particularly the $f_1(1285)$ and $f_1(1420)$ resonances, has been a topic of longstanding debates in hadron physics due to their unique structural characteristics. In theoretical calculations, the $f_1(1285)$ and $f_1(1420)$, which predominantly decay via $K\bar{K}^*+c.c.$, $\rm \pi\pi\eta$ and 4$\pi$ channels~\cite{PDG2024}, exhibit complex quark-gluon dynamics. One current explanation suggests that isospin-singlet ($1^3P_1$) states may form a mass eigenstate mixture under flavor SU(3) symmetry, with $f_1(1420)$ dominated by $s\bar{s}$ and $f_1(1285)$ by $u\bar{u}$ and $d\bar{d}$ components, parametrized by the mixing angle $\theta_{f_1}$~\cite{Cheng2011pb}.
There are also other explanations. Reference~\cite{950340152017} claims that $f_1(1420)$ is not a genuine resonance and it shows up as the manifestation of $K^*\bar{K}+c.c.$ and $\pi a_0(980)$ decay modes of  $f_1(1285)$ around 1420 MeV$/c^2$ due to a triangle singularity. Another study in Ref.~\cite{Xie_2020} assumes that the $f_1(1285)$ is a dynamically generated state from the strong $\bar{K^*}K+c.c.$ interaction, and in this picture the $f_1(1285)$ state has a strong coupling to the $\bar{K}K^*+c.c.$ channel. 

\begin{figure}[htp]
\begin{center}
\begin{tikzpicture}[scale=0.9]
\coordinate[label=below:$\bar s$]       (i1) at (0.0, 0.0);
\coordinate[label=above:$c$]            (i2) at (0.0, 1.0);
\coordinate[label=below:$\bar s$]       (o1) at (6.0, 0.0);
\coordinate[label=above:$s$]            (o2) at (6.0, 1.0);
\coordinate[label=right:$\nu_e$]        (o3) at (6.0, 1.7);
\coordinate[label=right:$e^+$]          (o4) at (6.0, 3.0);
\coordinate[label=120:$V_{cs}$]         (v1) at (3.0, 1.0);
\coordinate (v2) at (4.5, 2.0);
\coordinate[label=left:$D_s^+$] (D00) at (-0.5, 0.5);
\coordinate (D01) at (0, 0.5);
\coordinate[label=right:$f_1$] (H00) at (6.5, 0.5);
\coordinate (H01) at (6, 0.5);

\draw[anti-fermion] (i1) -- (o1);
\draw[fermion] (i2) -- (v1);
\draw[fermion] (v1) -- (o2);
\draw[photon, thick] (v1) -- node[label=above:$W^+$] {} (v2);
\draw[fermion] (v2) -- (o3);
\draw[anti-fermion] (v2) -- (o4);

\fill[cyan!70] (v1) circle (0.12);
\fill[blue!50] (0,0.5) ellipse (0.15 and 0.5);
\fill[green!60] (6,0.5) ellipse (0.15 and 0.5);

\end{tikzpicture}
\caption{Feynman diagram of $D^+_s \to f_1 e^+\nu_e$. }
\label{feynman}
\end{center}
\end{figure}
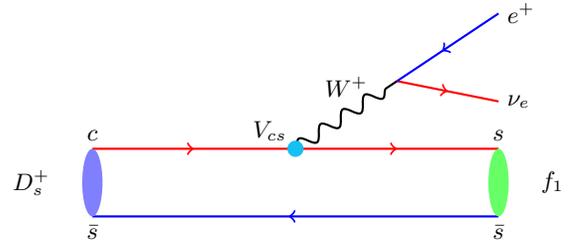

The SL decays $D_s^+\to f_1e^{+} \nu_e$ (where $f_1$ denotes $ f_1(1285)$ or $f_1(1420)$) can proceed via the Feynman diagram as shown in Fig.~\ref{feynman}. The charge conjugate channels are always implied throughout this paper. References~\cite{Cheng2017pcq,Qiao_2024,1110930012025} have predicted the branching fraction (BF) of $D_s^+\to f_1e^{+} \nu_e$ to be at the level of $10^{-4}$ based on the mixing picture between light and strange quarks . However, for SL $D_s^+$ decays into the axial-vector meson $f_1$ ,  there is no any experimental studies so far. Thus, experimental measurements of these two decay modes are important to test various theoretical calculations and to understand nonperturbative effects in heavy meson decays. Moreover, these measurements also can help people to understand the structure of the $f_1(1285)$ and $f_1(1420)$.

In this paper, we present the first search for $D^+_s\to\ f_1(1285)e^+\nu_e$ and $D^+_s\to\ f_1(1420)e^+\nu_e$ with $f_1(1285) \to\ \pi^+\pi^-\eta$ and $f_1(1420) \to\ K^+K^-\pi^0$, respectively. This analysis is performed by using $e^+e^-$ collision data corresponding to an integrated luminosity of 7.33 ~${\rm fb^{-1}}$ ~\cite{Ablikim_2015} collected at eight center-of-mass energy ($E_{\rm CM}$) points, as listed in Table~\ref{tab:mrec}, between 4.128 and 4.226 GeV with the BESIII detector.

\section{Experiment And Data Sample}
\label{sec:BESIII}

The BESIII detector~\cite{BESIII_2009fln} records symmetric $e^+e^-$ collisions provided by the BEPCII collider~\cite{Yu_2016cof}
in the center-of-mass energy range from 1.84 to 4.95 GeV,
with a peak luminosity of $1.1 \times 10^{33}\;\text{cm}^{-2}\text{s}^{-1}$ achieved at $\sqrt{s} = 3.773\;\text{GeV}$.
BESIII has collected large data samples in this energy region~\cite{Ablikim_202044040001,344_3372020}. The cylindrical core of the BESIII detector covers 93\% of the full solid angle and consists of a helium-based multilayer drift chamber~(MDC), a time-of-flight system~(TOF), and a CsI(Tl) electromagnetic calorimeter~(EMC), which are all enclosed in a superconducting solenoidal magnet providing a 1.0~T magnetic field.
The solenoid is supported by an octagonal flux-return yoke with resistive plate counter muon identification modules interleaved with steel.
The charged-particle momentum resolution at $1~{\rm GeV}/c$ is $0.5\%$, and the ${\rm d}E/{\rm d}x$ resolution is $6\%$ for electrons
from Bhabha scattering. The EMC measures photon energies with a resolution of $2.5\%$ ($5\%$) at $1$~GeV in the barrel (end-cap)
region. The time resolution in the plastic scintillator TOF barrel region is 68~ps, while that in the end-cap region was 110~ps.
The end-cap TOF system was upgraded in 2015 using multigap resistive plate chamber technology, providing a time resolution of 60~ps~\cite{Li_2017jpg,Guo_2017sjt,Ablikim_1130022022},
which benefits 83\% of the data used in this analysis.

Monte Carlo (MC) simulated samples produced with a {\sc geant4}-based~\cite{GEANT4_2002zbu} software package, which
includes the geometric description of the BESIII detector and the detector response, are used to determine the detection efficiencies
and to estimate the backgrounds. The simulation models the beam energy spread and initial state radiation (ISR) in the $e^+e^-$
annihilations with the generator {\sc kkmc}~\cite{Jadach_2000ir,Jadach_1999vf}.
The known decay modes are modeled with {\sc evtgen}~\cite{Lange_2001uf,Ping_2008zz} using BFs taken from the Particle Data Group~(PDG)~\cite{PDG2024},
and the remaining unknown decays from the charmonium states with {\sc lundcharm}~\cite{Chen_2000tv,Yang_2014vra}.
The final state radiation~(FSR) from charged final-state particles are incorporated using the {\sc photos} package~\cite{Richter_Was_1992hxq}.
Inclusive MC samples with an equivalent luminosity of 40 times that of the data are employed.
The samples contain all known open charm decay processes, the $c\bar{c}$ resonances, $J/\psi$, $\psi(3686)$, and $\psi(3770)$ via the ISR,
the hadronic continuum processes ($e^+e^-\to q\bar{q},q=u, d, {\rm and}~s$), Bhabha scattering, $\mu^+\mu^-$, $\tau^+\tau^-$, and two-photon processes.

The signal decays $D^+_s\to\ f_1(1285)e^+\nu_e$ and $D^+_s\to\ f_1(1420)e^+\nu_e$ are simulated using the ISGW2 model~\cite{5227831995}. The $f_1(1285)$ decays into $a_0(980)^-(\to\pi^-\eta)\pi^+$, $a_0(980)^+(\to\pi^+\eta)\pi^-$ and $\pi^+\pi^-\eta$ final sates~\cite{PDG2024}.
The $f_1(1420)$ decays into $K^{*+}(\to K^+\pi^0)K^-$, $K^{*-}(\to K^-\pi^0)K^+$ and $a_0(980)^0(\to K^+K^-)\pi^0$ final states\cite{1040320112021}.
A relativistic Breit-Wigner function is used to parametrize the resonances $f_1(1285)$ and $f_1(1420)$, whose masses and widths are fixed to the PDG values~\cite{PDG2024}.

\section{DATA ANALYSIS}
\label{sec:selection}
\bigskip

In $e^+e^-$ collision data taken at center-of-mass energies between 4.128 and 4.226 GeV, the $D^{\pm}_s$ mesons are mostly produced in the process $e^+e^-\to D^{*\pm}_sD^{\mp}_s$,
where $D^{*\pm}_s$ decays to $\gamma D^{\pm}_s$ with a BF of $(93.6\pm0.4)$\%~\cite{PDG2024}.
This allows to measure the BF of an SL decay using a double-tag (DT) technique as employed in the analysis~\cite{BESIII_2018cfe}.
First, we reconstruct a $D^-_s$ meson in a hadronic decay mode which is denoted as a ``single-tag (ST)" candidate.
Then, in the rest of the event recoiling against the tagged $D^-_s$ meson,
we require a positron and an $f_1$ from the signal SL decay
in the presence of one additional transition photon from the $D^{*\pm}_s$ decay. Such an event is denoted as a DT candidate.
With a reconstructed ST sample and its DT sub-sample, the absolute BF is determined by
\begin{equation}
{\cal B}_{\rm SL}=\frac{N^{\rm obs}_{\rm DT}}{{\cal B}_{\rm sub}{\cal B}_{\gamma  }\sum_{\alpha} N^{\alpha}_{\rm ST}\epsilon^{\alpha}_{\rm DT}/\epsilon^{\alpha}_{\rm ST}},
\label{eq:Bsig-gen}
\end{equation}
where ${\cal B}_{\gamma}$ and ${\cal B}_{\rm sub}$ are the BFs of $D^{*\pm}_s \to\gamma D^{\pm}_s$ and $\eta\to\gamma\gamma$/$\pi^0\to\gamma\gamma$, respectively.
$N^{\rm obs}_{\rm DT}$ is the total DT signal yield with all tag modes.
For the tag mode $\alpha$, $N^{\alpha}_{\rm ST}$ is the observed ST yield,
$\epsilon^{\alpha}_{\rm DT}$ and $\epsilon^{\alpha}_{\rm ST}$ are the DT and ST efficiencies, respectively,
which can be obtained from the signal and the inclusive MC samples.

\subsection{Selection of ST events}
\label{subsec:ST}
\bigskip
In this analysis, for the decay $D^+_s\to\ f_1(1420)e^+\nu_e$, the ST $D_s^-$ mesons are reconstructed with 9 hadronic decay modes:
$D^-_s\to K^{+}K^{-}\pi^{-}$,
$D^-_s\to K^0_{S}K^{-}$,
$D^-_s\to \pi^{-}\eta$,
$D^-_s\to \pi^{-}\eta^{\prime}_{\pi^{+}\pi^{-}\eta}$,
$D^-_s\to K^{+}K^{-}\pi^{-}\pi^{0}$,
$D^-_s\to K^0_{S}K^{+}\pi^-\pi^-$,
$D^-_s\to \rho^{-}\eta$,
$D^-_s\to \pi^{-}\eta^{\prime}_{\gamma\rho^0}$,
and $D^-_s\to K^{0}_{S}K^{-}\pi^{0}$.
For the decay $D^+_s\to\ f_1(1285)e^+\nu_e$, the ST $D_s^-$ mesons are reconstructed by 12 hadronic decay modes, including three additional tag modes of
$D^-_s\to \pi^{+}\pi^{-}\pi^{-}$,
$D^-_s\to K^{-}\pi^{+}\pi^{-}$,
and $D^-_s\to K^0_{S}K^{-}\pi^+\pi^-$. The configuration of these two ST combinations is guided by the optimization of the figure of merit $\epsilon/(\frac{3}{2}+\sqrt{B})$, which excludes the hadron decay modes with low signal efficiency and high background and thereby helps improve the signal significance for the decay  $D^+_s\to\ f_1 e^+\nu_e$  . Here, $\epsilon$ and $B$  denote the signal efficiency estimated by the signal MC samples and the background yields estimated by the inclusive MC samples, respectively.

For a given ST mode, the invariant mass of the corresponding $D^-_s$ candidates are reconstructed
by all possible combinations of selected $K^{\pm}$, $\pi^{\pm}$, $K^0_S$, $\pi^0$, $\eta$, and $\eta^{\prime}$ candidates in the tagged event.

Charged tracks, except for the daughter tracks of $K^0_S$ candidates,
are selected by requiring its closest approach to the interaction point (IP)
within $\pm10$ cm along the beam direction and within 1 cm in the plane perpendicular to the beam direction.
In addition, the polar angle ($\theta$) of the charged track relative to the beam direction must be
within the detector acceptance by requiring $|\!\cos\theta| \le 0.93$.
Charged particle identification (PID) is performed by combining measurements of the
energy deposited in the MDC (${\rm d}E/{\rm d}x$) and the flight time in the TOF.
Tracks are identified by the PID likelihood $\mathcal{L}_h$ ($h=\pi,\ K$) for each hadron $h$ hypothesis.
A pion (kaon) candidate is required to satisfy $\mathcal{L}_{\pi(K)}>\mathcal{L}_{K(\pi)}$ and $\mathcal{L}_{\pi(K)}>0$.

Photon candidates are selected from isolated electromagnetic showers
which have a minimum energy of 25 MeV in the EMC barrel region ($|\!\cos\theta|< 0.80$)
or 50 MeV in the EMC end-cap region ($0.86 < |\!\cos\theta| < 0.92$).
To reduce the number of photon candidates that result from noise and beam backgrounds, 
the difference between the EMC time and the event start time is required to be less than 700 ns after the beam collision.
The opening angle between a photon candidate and the closest charged track is required to be greater than 10$^\circ$ to exclude showers that originate from charged tracks.

The $\pi^0$ and $\eta$ candidates are reconstructed from pairs of photon candidates
that have an invariant mass within the intervals (0.115, 0.150) and (0.490, 0.580) GeV/$c^2$, respectively.
To improve the momentum resolution, a kinematic fit is performed,
constraining the $\gamma\gamma$ invariant mass to the known mass of $\pi^0$ or $\eta$~\cite{PDG2024}, and the $\chi^2$ of the kinematic fit is required to be less than 30 to reject the combinatorial background.

The $K^0_S$ candidates are reconstructed from two oppositely charged tracks without imposing PID criteria,
and the final-state tracks are required to have a trajectory that approaches the IP to within $\pm20$ cm
along the beam direction and to have $|\!\cos\theta| <0.93$.
A vertex-constrained fit is performed to all oppositely charged track pairs of $K^0_S$ candidates,
and the resulting track parameters are used to obtain the invariant masses
which are required to be within the interval (0.487, 0.511) GeV/$c^2$.
To further suppress the combinatorial background, the $\chi^2$ of the vertex fit is required to be less than 200.

The $\eta^{\prime}$ candidates are reconstructed via the decay modes $\gamma\rho^0$ and $\pi^+\pi^-\eta$
by requiring the corresponding invariant masses to be within the intervals (0.938, 0.978) and (0.943, 0.973) GeV/$c^2$, respectively.
Additionally, the minimum energy of the $\gamma$ from $\eta^{\prime}\to\gamma\rho^0$ is required to be greater than 100 MeV.
The $\rho^0$ candidates are reconstructed from $\pi^+\pi^-$ pairs within invariant mass interval of (0.570, 0.970) GeV/$c^2$.
The $\rho^-$ candidates are reconstructed from $\pi^0\pi^-$ combinations within invariant mass interval of (0.625, 0.925) GeV/$c^2$.

To suppress the backgrounds from $D^{*\pm}$ decays,
the momentum of any pion not originating from a $K^0_S$ decay is required to be greater than 100 MeV/$c$.
Since both ST modes $D^-_s\to K^0_SK^-$ and $D^-_s\to K^-\pi^+\pi^-$ are used,
we require the $\pi^+\pi^-$ invariant mass from $D^-_s\to K^-\pi^+\pi^-$ to be outside of the range (0.487, 0.511) GeV/$c^2$
to exclude the $K^0_S$ pollution from the process $D^-_s\to K^0_SK^-$.

To further suppress the non-$D^{*\pm}_sD^{\mp}_s$ backgrounds, a variable that represents the
invariant mass of the system recoiling against the selected $D^-_s$ candidate is defined as
\begin{equation}
\begin{aligned}
&m_{\rm rec}=\\
&\sqrt{\frac{(E_{\rm CM}-\sqrt{\bm{p}^2_{D^-_s}c^2+m^2_{D_s}c^4})^2}{c^4}-\frac{(\bm{p}_{\rm CM}-\bm{p}_{D^-_s})^{2}}{c^2}},
\end{aligned}
\label{def:Mrec}
\end{equation}
where $E_{\rm CM}$ ($\bm{p}_{\rm CM}$) is the center-of-mass energy (momentum) of $e^+e^-$ collision, $\bm{p}_{D^-_s}$ is the measured $D^-_s$ momentum in the center-of-mass system, and $m_{D_s}$ is the known $D_s$ mass~\cite{PDG2024}. In the process $e^+e^-\to D^{*\pm}_sD^{\mp}_s$, the selected $D^-_s$ candidates are produced either directly from $e^+e^-$ collision or from the decay $D^{*\pm}_s\to\gamma D^{\pm}_s$. The corresponding $m_{\rm rec}$ distribution for the former case peaks at the known $D^{*\pm}_s$ mass~\cite{PDG2024} smeared by the mass resolution, and the distribution for the latter case is relatively flat within about $\pm65\mevcc$ around the former peak. Based on this, the $D^-_s$ candidates are accepted within the regions shown in Table~\ref{tab:mrec}. For an event with multiple candidates for a specific tag mode per charge, only the one with $m_{\rm rec}$ closest to the known $D^{*\pm}_s$ mass is chosen.

\begin{figure}[htp]
\begin{center}
\includegraphics[width=0.3\textwidth]{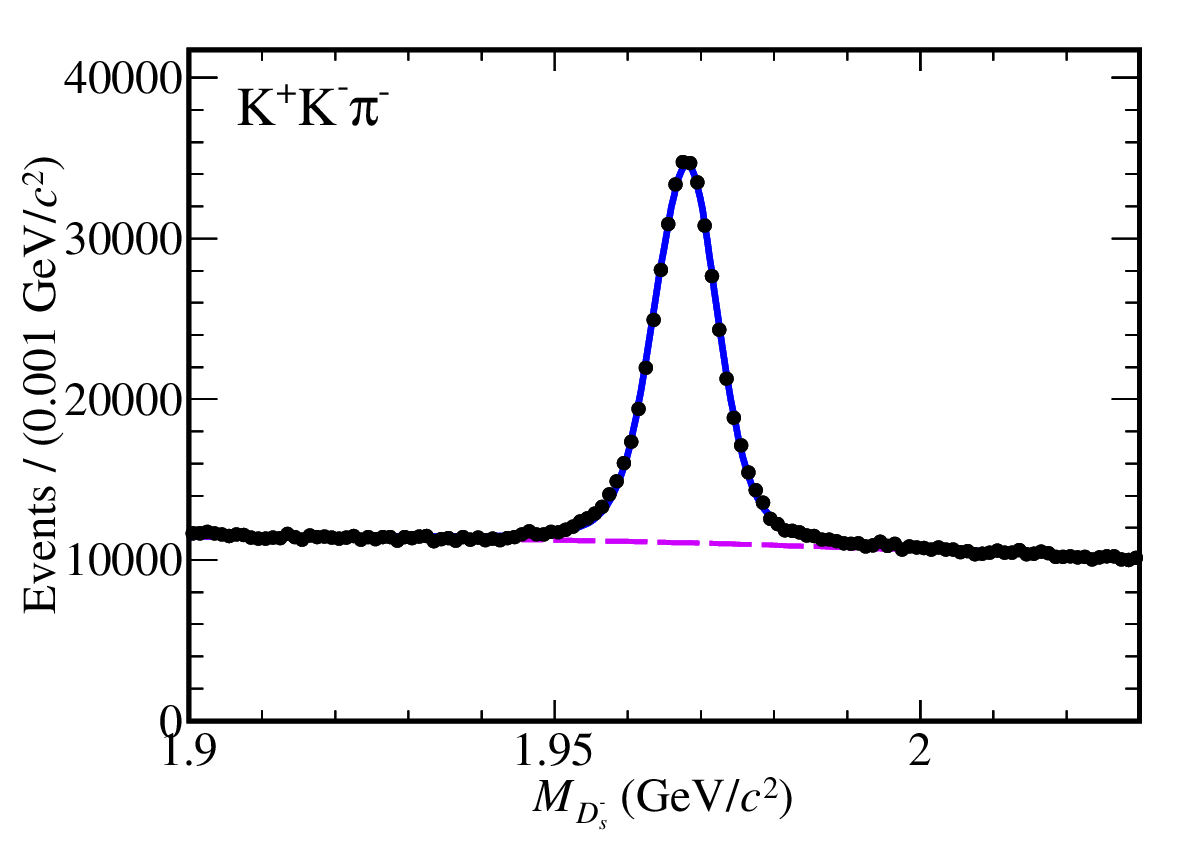}
\includegraphics[width=0.3\textwidth]{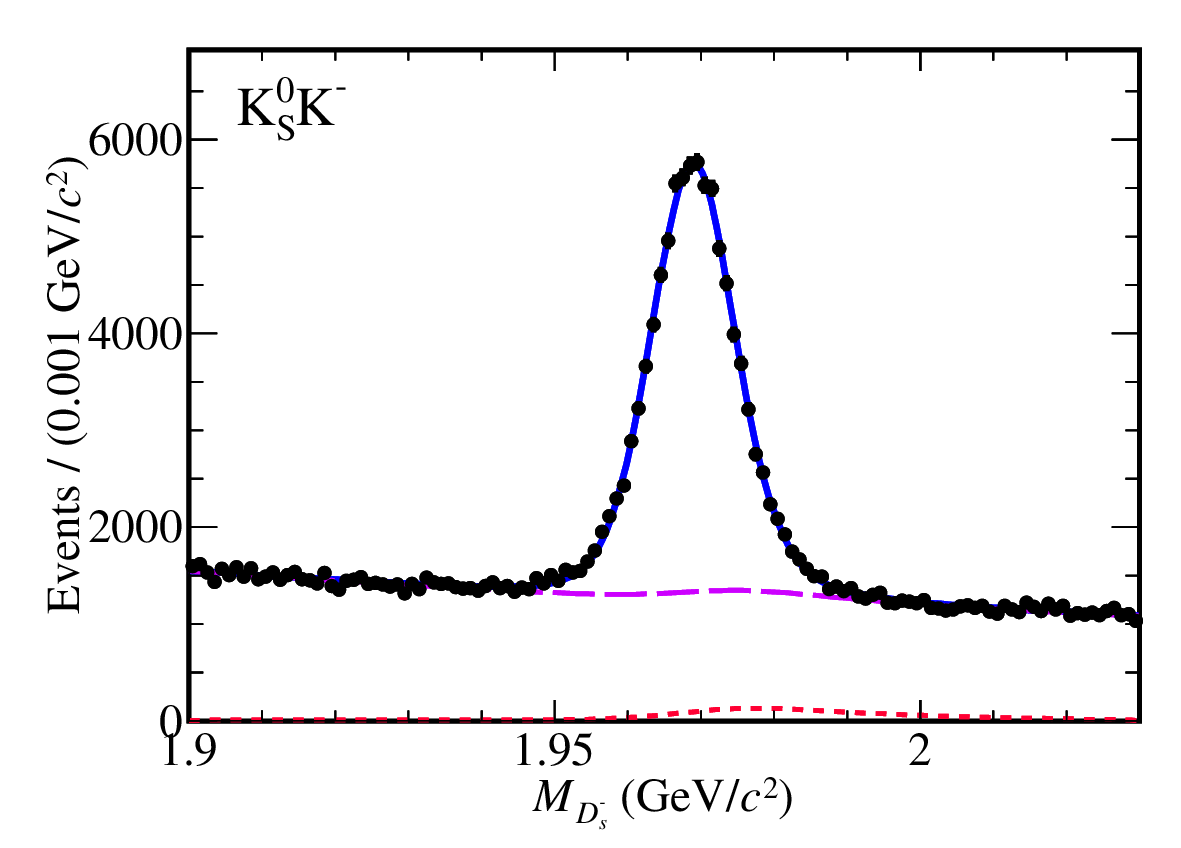}
\includegraphics[width=0.3\textwidth]{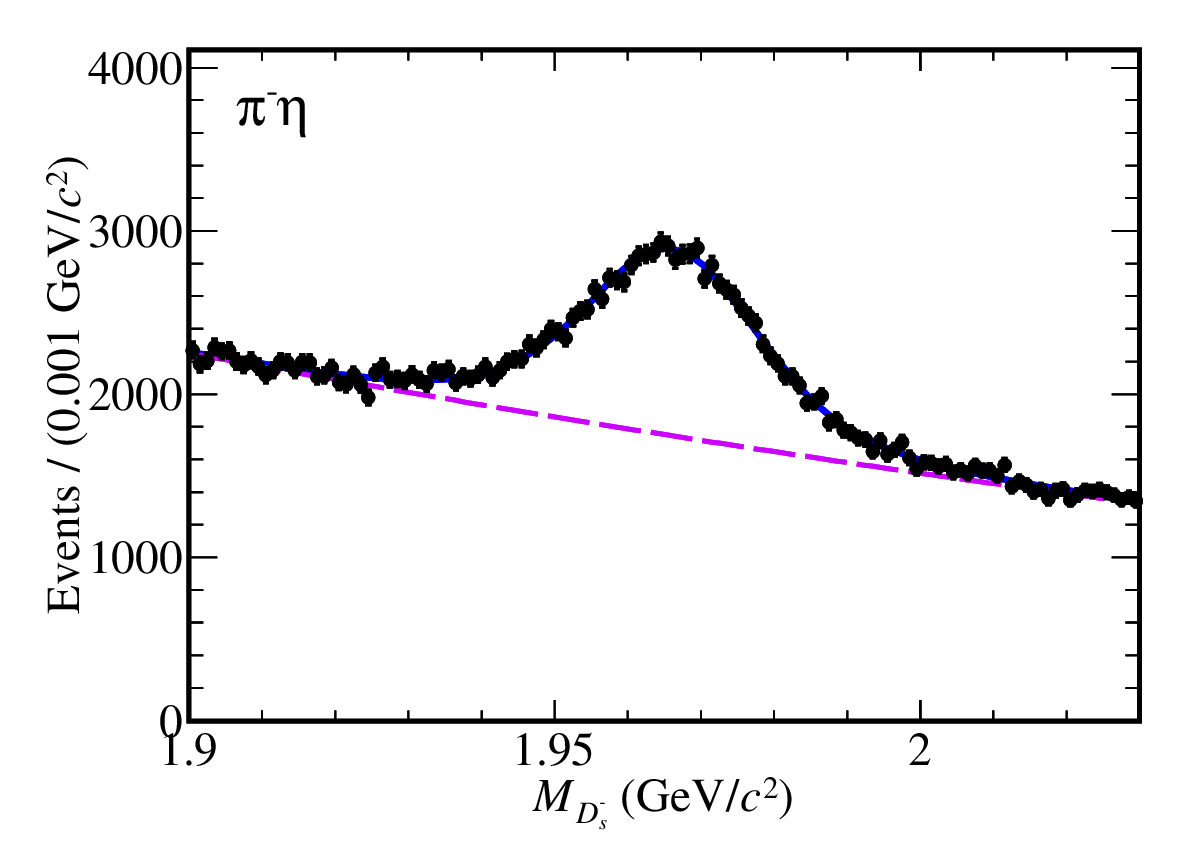}
\includegraphics[width=0.3\textwidth]{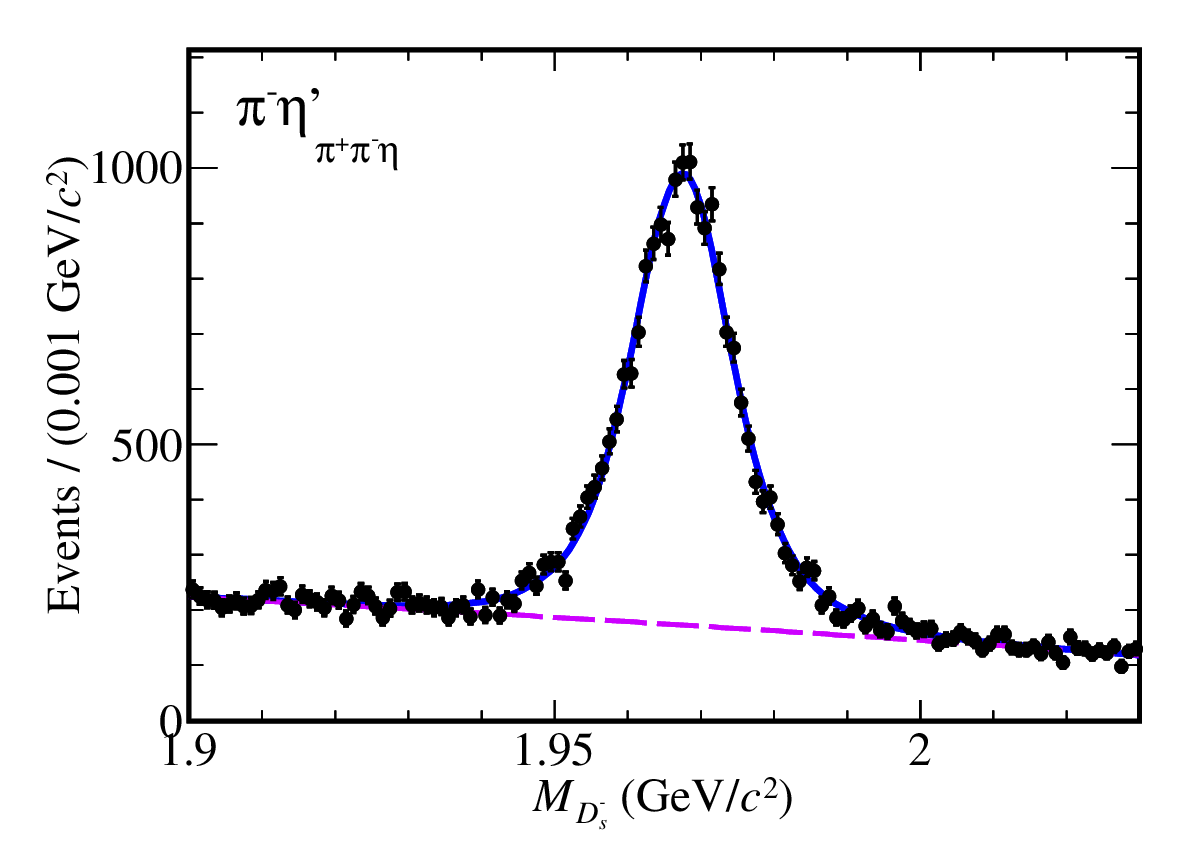}
\includegraphics[width=0.3\textwidth]{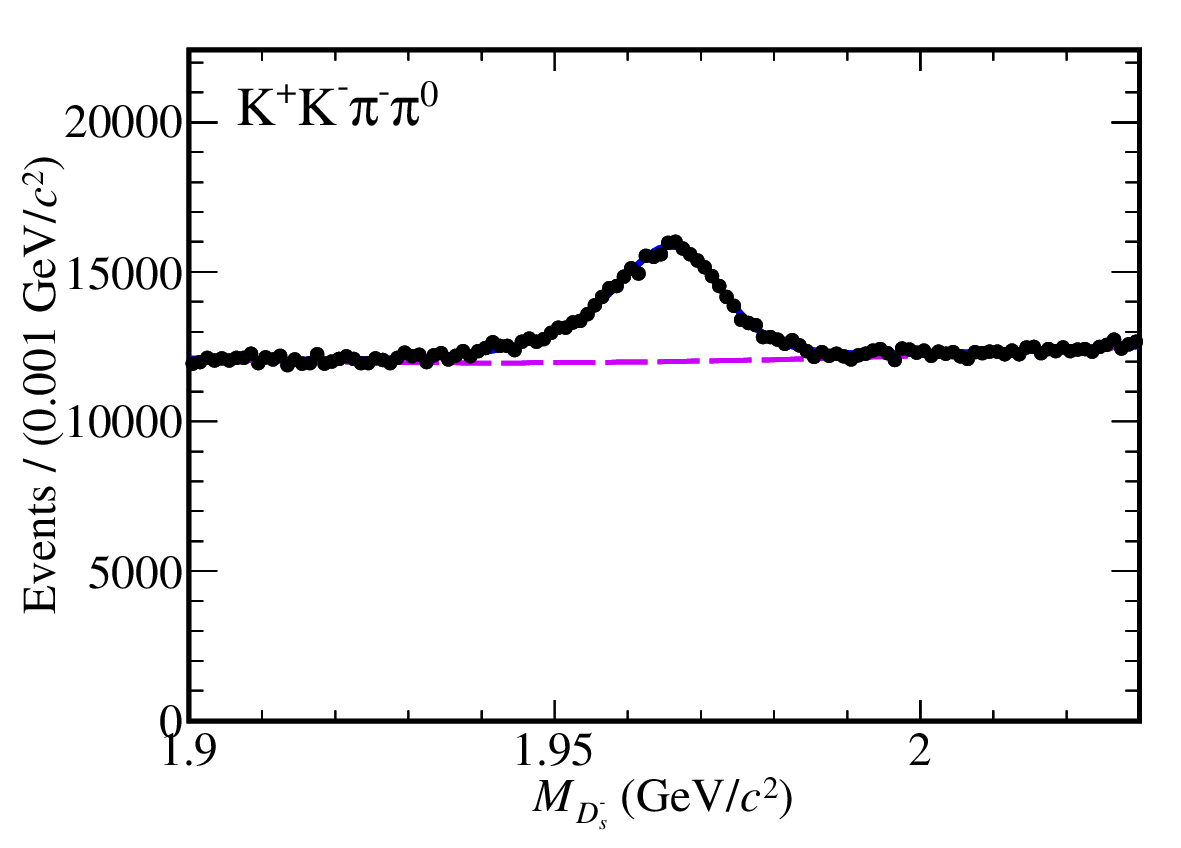}
\includegraphics[width=0.3\textwidth]{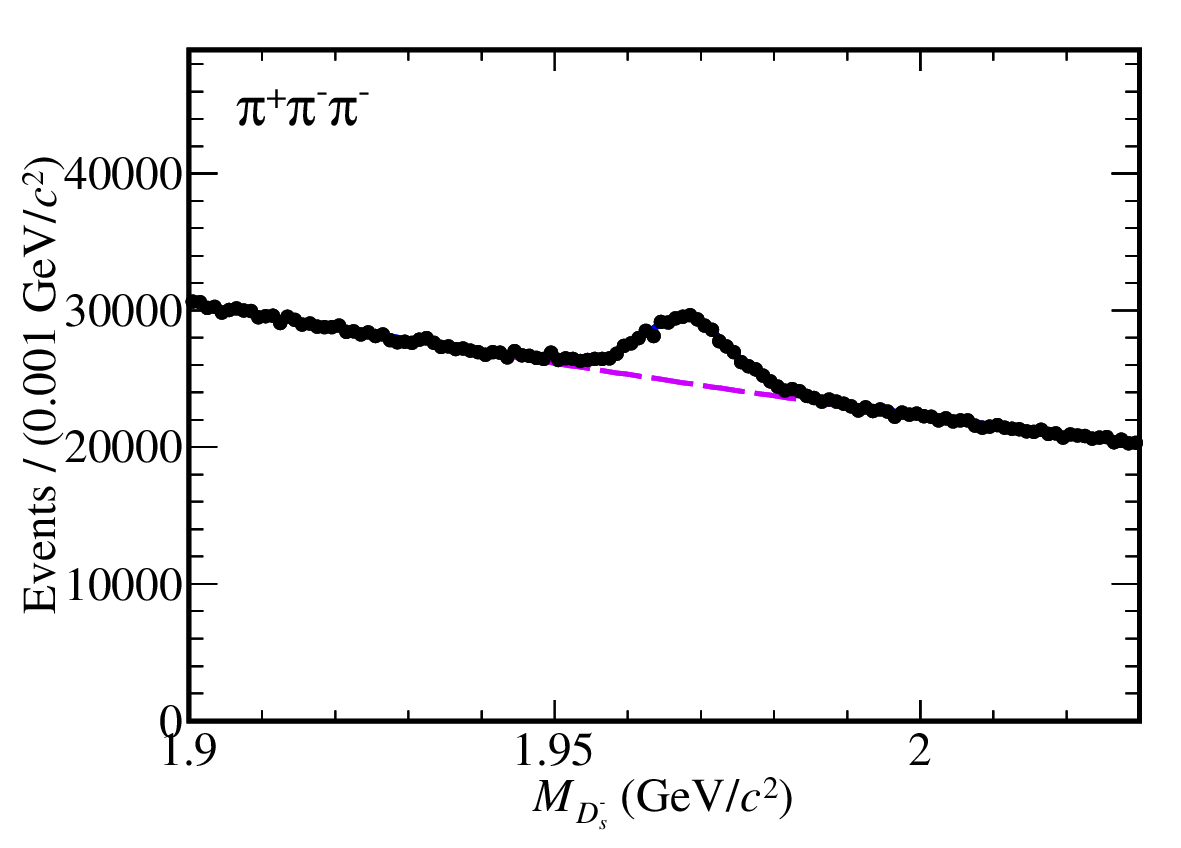}
\includegraphics[width=0.3\textwidth]{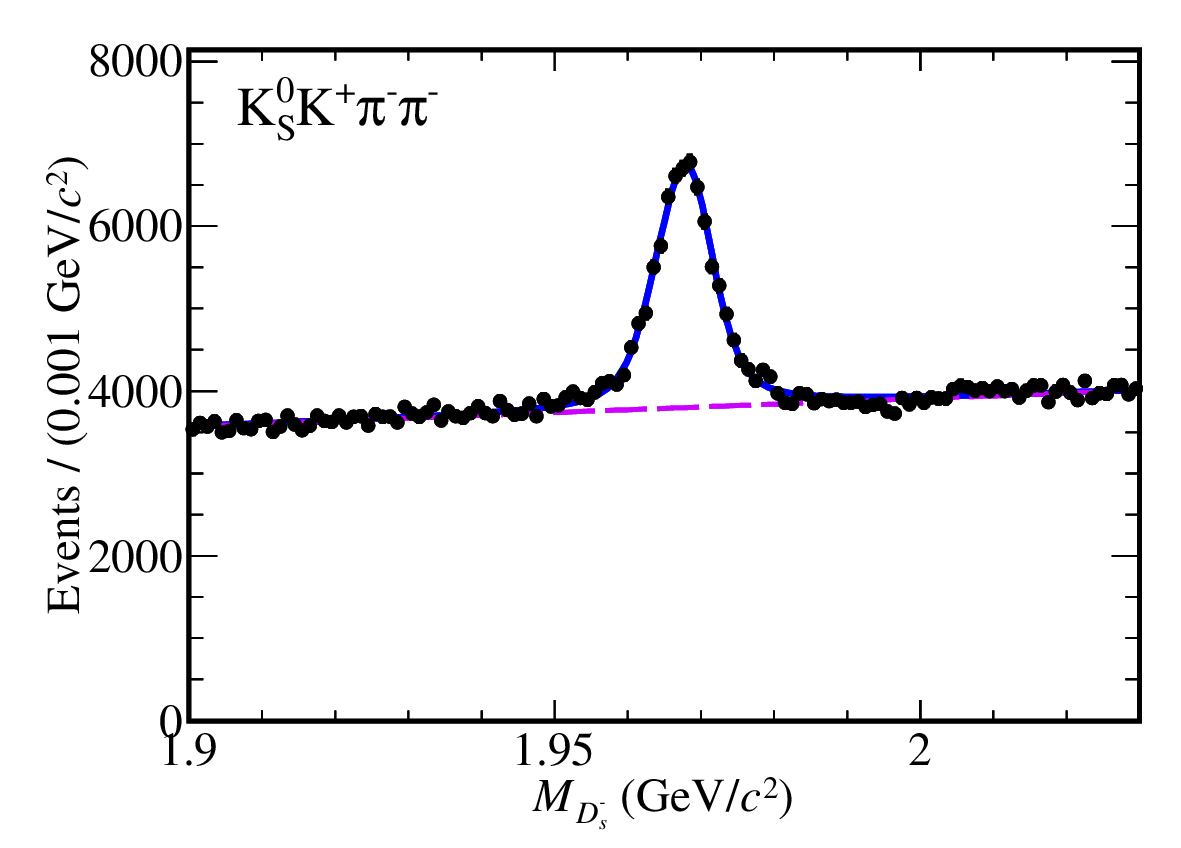}
\includegraphics[width=0.3\textwidth]{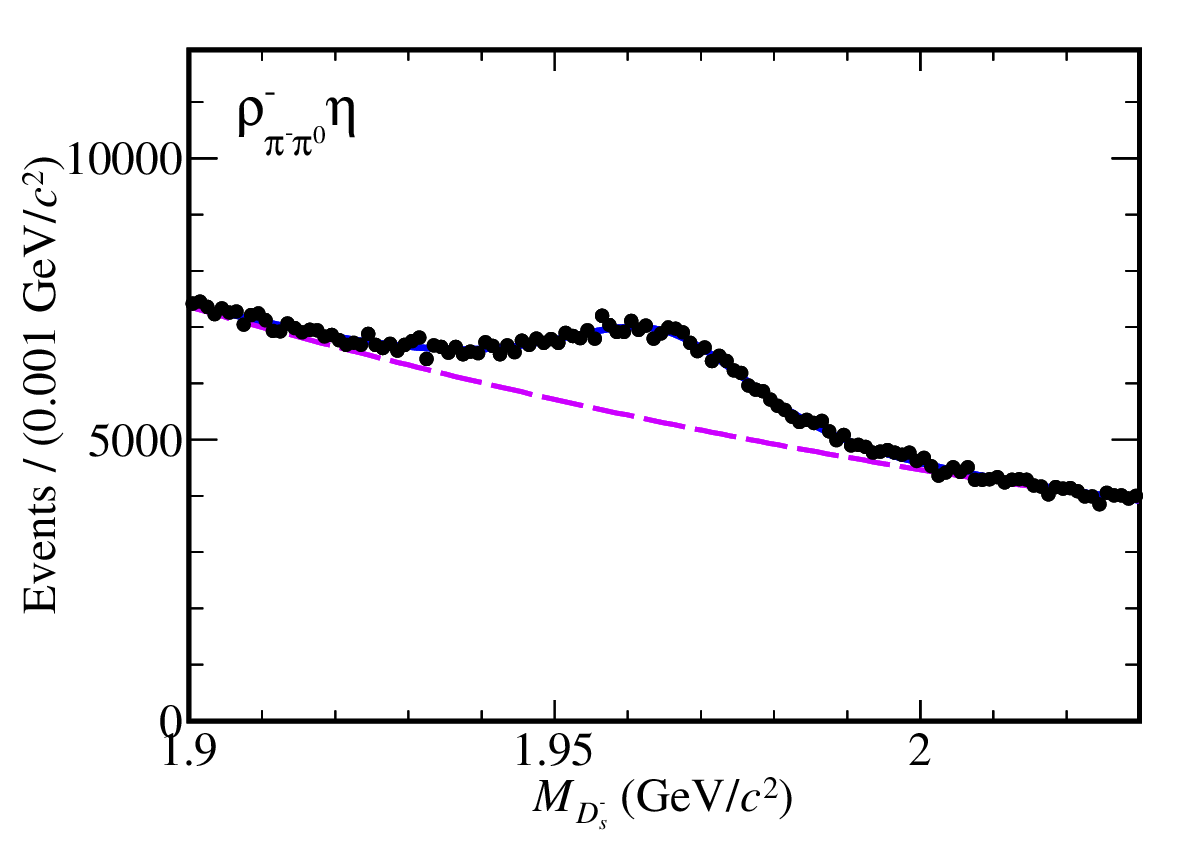}
\includegraphics[width=0.3\textwidth]{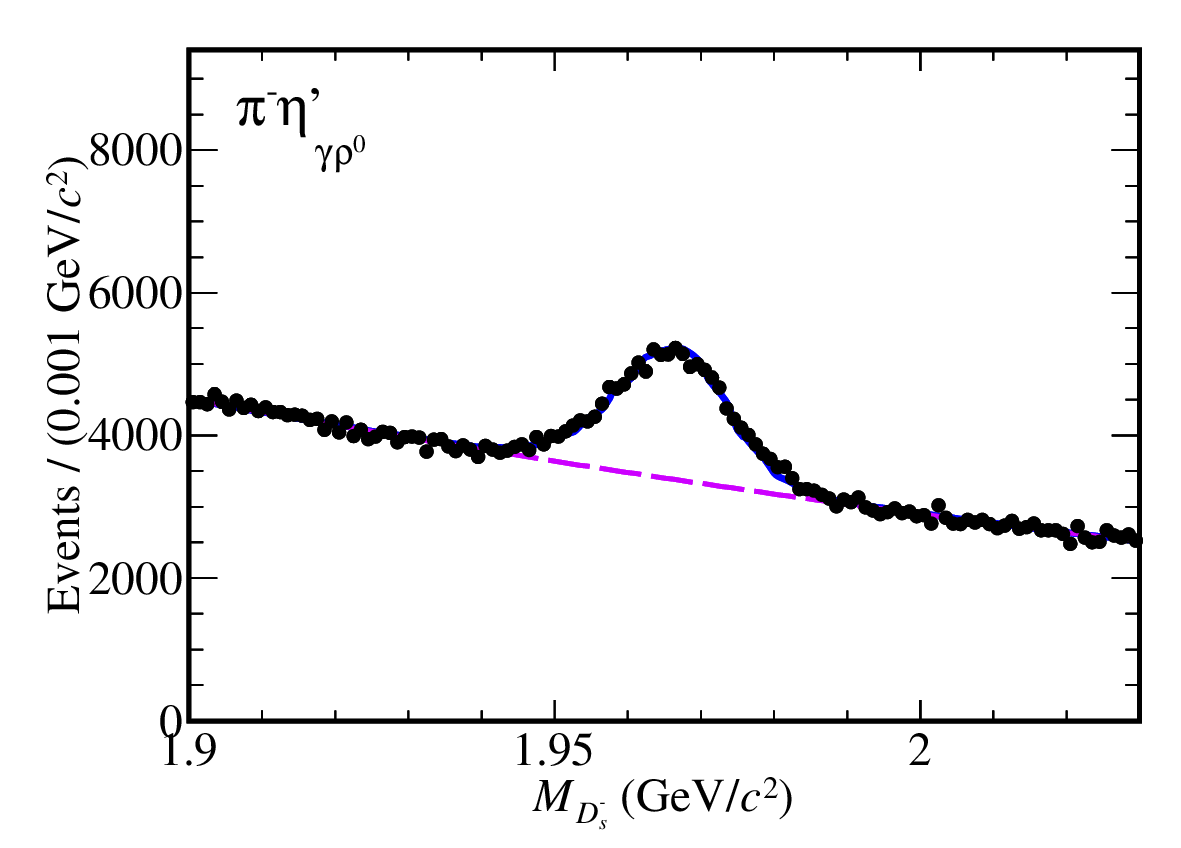}
\includegraphics[width=0.3\textwidth]{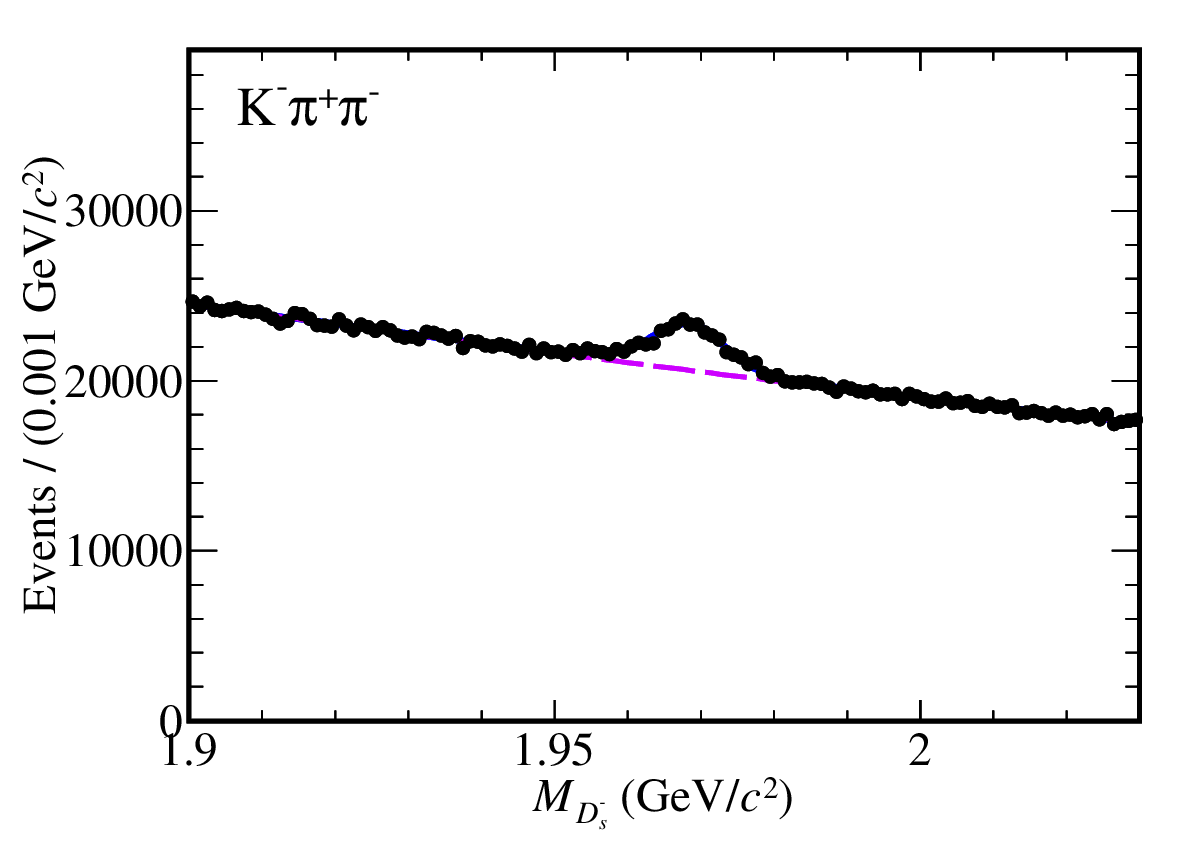}
\includegraphics[width=0.3\textwidth]{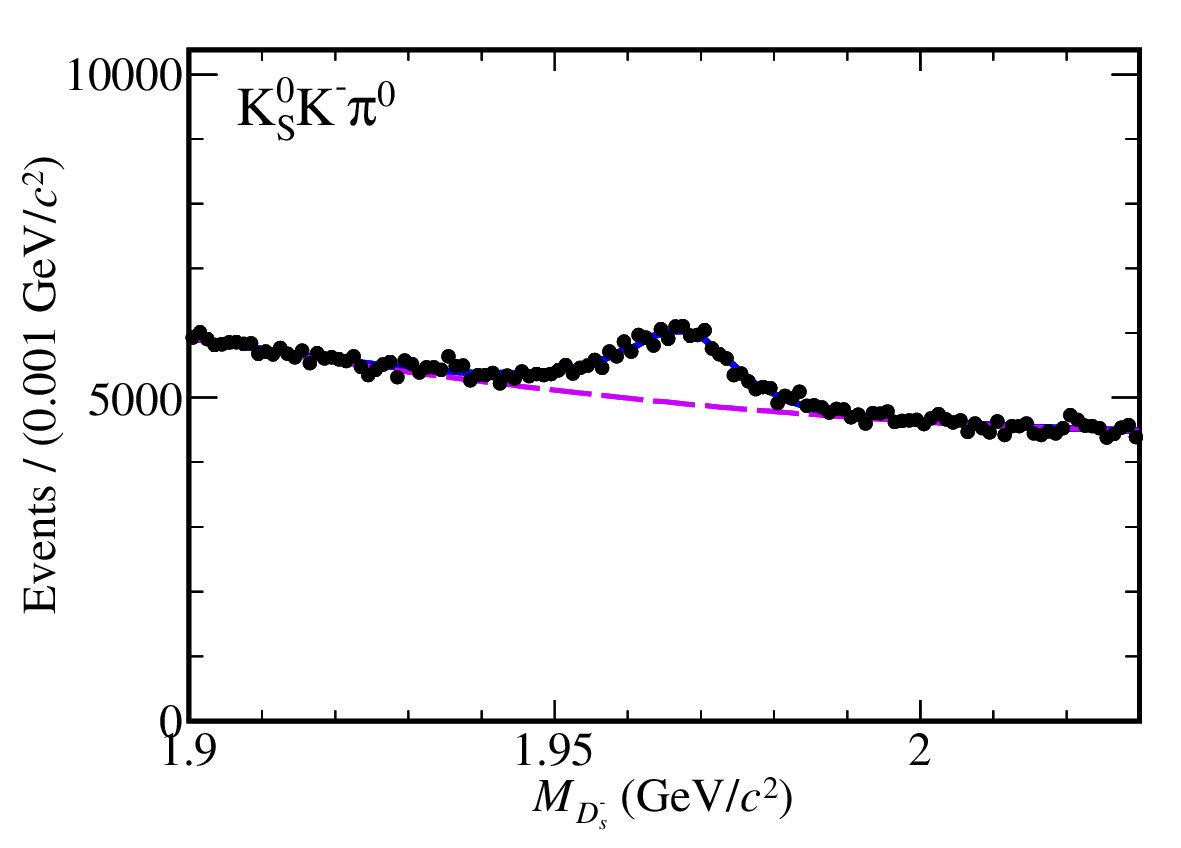}
\includegraphics[width=0.3\textwidth]{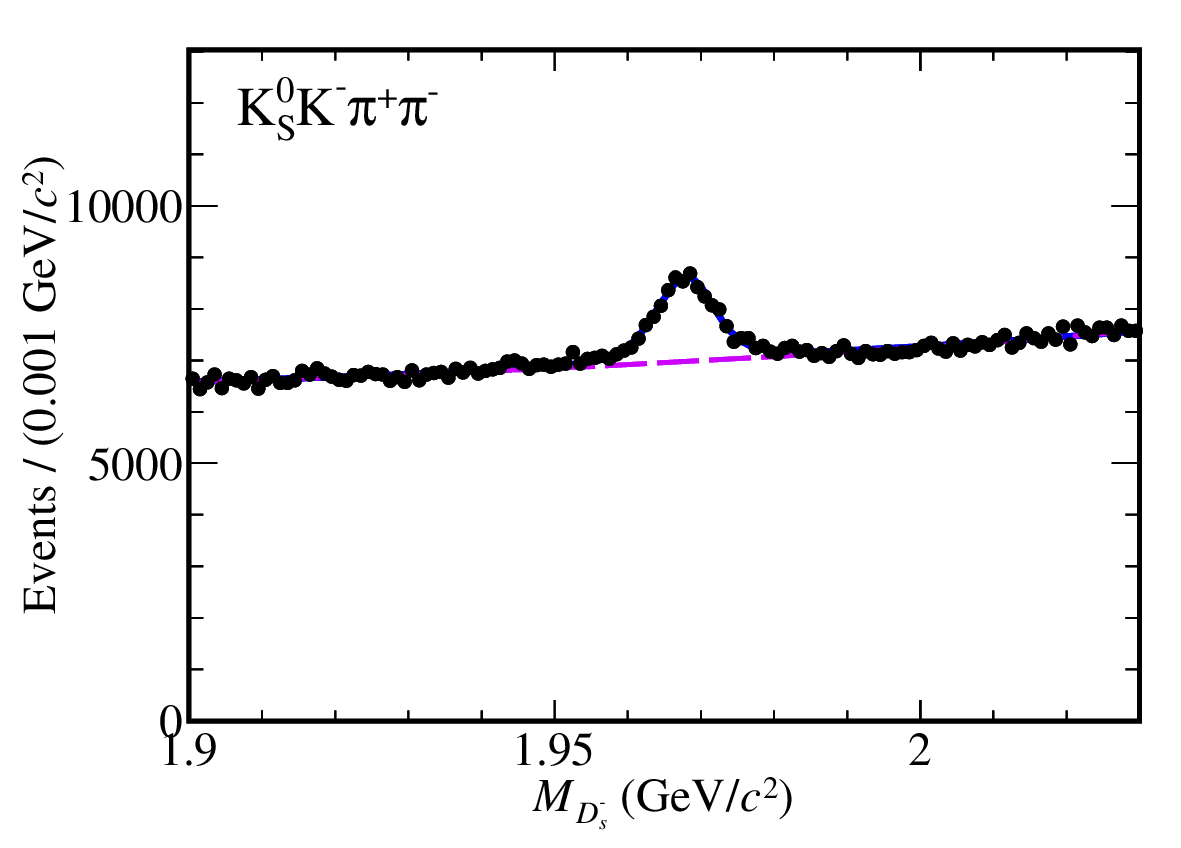}
\caption{Fits to the tag ${M}_{D_{s}^{-}}$ distributions of data samples at 4.128$\sim$4.226 GeV. 
The points with error bars are data.
The blue solid curves are the total fits. 
The violet long-dashed curves are the fitted backgrounds,
and the red dotted curve in the $K^0_{S}K^{-}$ mode is $D^-\to K^0_{S}\pi^{-}$ background.}
\label{ST_fit}
\end{center}
\end{figure}

The distributions of the tag $D^-_s$ mass (\(M_{\rm tag}\)) for events
satisfying the above criteria, with the  binned maximum
likelihood fits superimposed, are shown in Fig.~\ref{ST_fit} for the data samples of
$E_{\rm cm}$ = 4.128$\sim$4.226 GeV. In each fit, the signal is described by the MC-simulated lineshape convolved with a Gaussian function accounting for the resolution difference between the data and MC simulation, where the parameters of the Gaussian are free to vary.
The non-peaking background is modeled by a first-order Chebyshev polynomial.
For the tag mode $D^-_s\to K^0_SK^-$, the peaking background from $D^-\to K^0_S\pi^-$ is described by the simulated MC shape that is smeared
with the same Gaussian function as in the signal shape, with its yield determined from the fit.
The ST yield is determined from the fit for each ST mode. Then, the events within the $D^-_s$ mass regions are selected for further analysis.
The requirements on the $D^-_s$ mass regions, the ST yields, and the corresponding ST detection efficiencies obtained
with the inclusive MC samples for various tag modes, are the same as those in Ref.~\cite{1321419012024}. Summing over various tag modes and energy points, we obtain the total ST yield $N_{\rm tag}^{\rm tot} = (771.1 \pm 3.4)\times10^3$.

\begin{table}[htp]
\begin{center}
\caption{The integrated luminosities and the requirements of $m_{\rm rec}$ for each data set.}
\vspace{0.50cm}
\begin{tabular}{l|c|c}
\hline
\hline
$E_{\rm CM}(\rm GeV)$ & Luminosity (pb$^{-1}$) & $m_{\rm rec}$(GeV/$c^2$)\\
\hline
4.128 &401.5  & [2.060, 2.150] \\
4.157 &408.7  & [2.054, 2.170] \\
4.178 &3189.0 & [2.048, 2.190] \\
4.189 &569.8  & [2.050, 2.180] \\
4.199 &526.0  & [2.046, 2.200] \\
4.209 &571.7  & [2.044, 2.210] \\
4.219 &568.7  & [2.042, 2.220] \\
4.226 &1091.7 & [2.040, 2.220] \\                   
\hline
\hline
\end{tabular}
\label{tab:mrec}
\end{center}
\end{table}

\subsection{Selection of DT events}
\label{subsec:DT}
\bigskip

The DT candidates are selected on top of the ST selection. At the recoiling side of the ST $D^-_s$ mesons,
the \Dsp\ SL candidates are reconstructed by requiring exactly three charged tracks identified as $\pi^+$, $\pi^-$, $e^+$ for the decay $D^+_s\to\ f_1(1285)e^+\nu_e$ and $K^+$, $K^-$, $e^+$ for the decay $D^+_s\to\ f_1(1420)e^+\nu_e$. In addition, as in the tag side, a reconstructed $\eta$ ($\pi^0$) is required for the decay $D^+_s\to\ f_1(1285)e^+\nu_e$ ($D^+_s\to\ f_1(1420)e^+\nu_e$).
Since most of $D^{*\pm}_s$ decay into the $\gamma D^{\pm}_s$ final state,
one extra good  photon candidate is required to reconstruct this transition.
To identify the kaons or pions from the \Dsp\ SL decay and the photon candidate from $D^{*\pm}_s$,
the same criteria as in the ST candidate selection are used.
Positron PID uses the measured information in the MDC, TOF and EMC.
The combined likelihoods ($\mathcal{L}^{\prime}$) under the positron, pion, and kaon hypotheses are obtained.
Positron candidates are required to satisfy $\mathcal{L}^{\prime}_e>0$ and $\mathcal{L}^{\prime}_e/(\mathcal{L}^{\prime}_e+\mathcal{L}^{\prime}_{\pi}+\mathcal{L}^{\prime}_K)>0.8$.
To reduce background from hadrons and muons, the positron candidate is further required to satisfy $E_{\rm EMC}/|\bm{p}_{e}|>0.8c$, where $E_{\rm EMC}$ is the deposited energy in the EMC, $\bm{p}_e$ is the three-momentum vector reconstructed in the MDC.

\onecolumngrid

\begin{figure}[htbp]
\begin{center}
\includegraphics[width=0.9\textwidth]{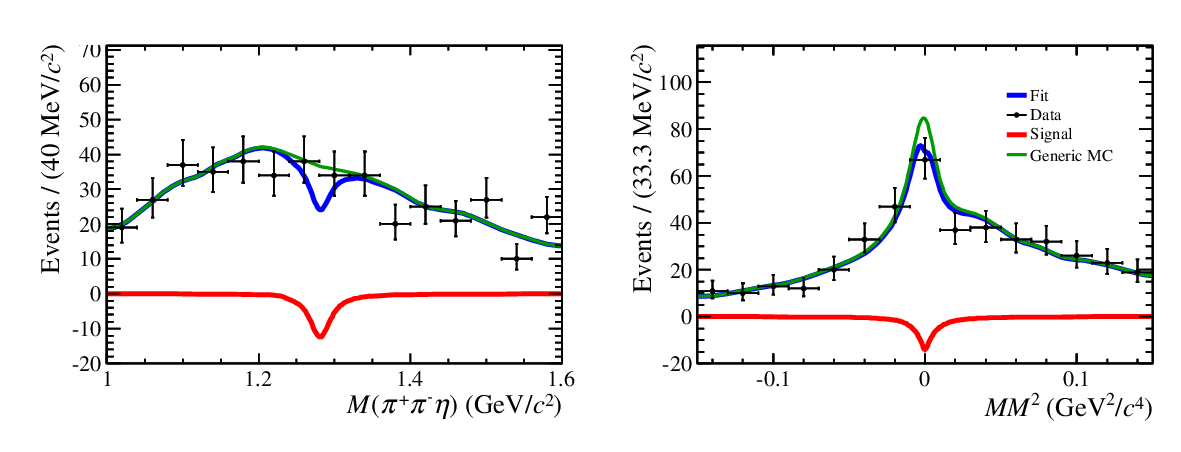}
\includegraphics[width=0.9\textwidth]{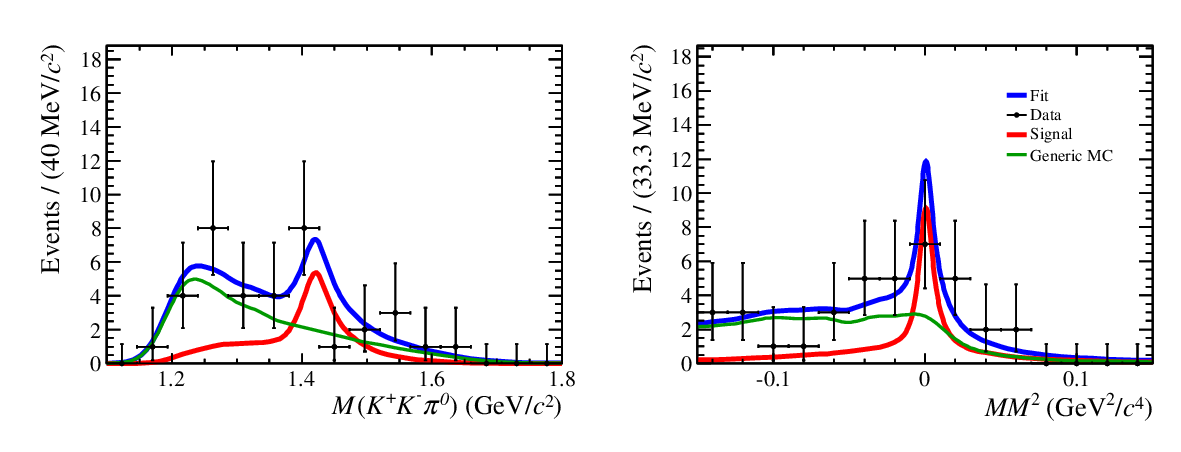}
\caption{
    Unbinned maximum likelihood fits to the two-dimensional $M_{\pi^{+}\pi^{-}\eta}$ and ${MM}^2$ distributions in the decay $D^{+}_{s} \to f_{1}(1285) e^{+}\nu_{e}$ (top) or $M_{K^{+}K^{-}\pi^{0}}$ and ${MM}^2$ distributions in the decay $D^+_s\to\ f_1(1420)e^+\nu_e$ (bottom).
The points with error bars are data, the blue lines are the fits, the red lines are the signals,
and the green lines are the backgrounds modeled based on the inclusive MC samples.
}
\label{2dfit:1420}
\end{center}
\end{figure}

\twocolumngrid

To choose the transition photon from multiple candidates, a kinematic
fit is performed to each event, requiring the total energy-momentum
conservation, and constraining the invariant mass of the ST \Dsm, the
signal \Dsp\ and the corresponding $D^{*\pm}_s$ to their known
masses~\cite{PDG2024}.  The four-vector of the undetected neutrino is
undetermined, so this is a three-constraint (3C) kinematic fit. The
kinematic fit is performed with two hypotheses, with one assuming the ST \Dsm\ as from the \DsSTm decay, and the other assuming \Dsp\ with SL decay as the daughter of the \DsSTp. Then the photon and the decay sequence in each event with the minimal fit $\chi^2_{\rm 3C}$ are selected.

Regardless of whether or not the photon forms a $D^{*-}_s$ with the ST $D^-_s$, for real $D^{*\pm}_sD^{\mp}_s$ events,
the recoiling mass squared ${M^2_{\rm rec}}$ against the photon and the ST $D^-_s$ should peak at $m^2_{D_s}$. 
It is calculated as
\begin{equation}
\begin{aligned}
&{M^2_{\rm rec}}=(E_{\rm CM}-E_{D^-_s}-E_{\gamma})^2/c^4\\
&-(\bm{p}_{\rm CM}-\bm{p}_{D^-_s}-\bm{p}_{\gamma})^2/c^2,
\end{aligned}
\label{def:M2rec}
\end{equation}
where $E_{D^-_s}$ ($\bm{p}_{D^-_s}$) is the energy (momentum) of the reconstructed ST \Dsm,
and $E_{\gamma}$ ($\bm{p}_{\gamma}$) is the energy (momentum) of the selected photon.
To improve the resolution, we constrain the invariant mass of the ST $D^-_s$ decay products to $m_{D_s}$.
To suppress the background,
we require $M^2_{\rm rec}$ to satisfy the optimized intervals (3.82, 4.05) GeV$^2/c^4$ for $D^+_s\to\ f_1(1285)e^+\nu_e$ and (3.78, 4.05) GeV$^2/c^4$ for $D^+_s\to\ f_1(1420)e^+\nu_e$.

To identify the SL decay with the information of the missing neutrino which is undetectable, we define
\begin{eqnarray}
\begin{aligned}
    & {MM}^2=(E_{\rm CM}-E_{D^-_s}-E_{\bm{h}}-E_{e}-E_{\gamma})^2/c^4 \\
&-(\bm{p}_{\rm CM} -\bm{p}_{D^-_s} -\bm{p}_{\bm{h}} -\bm{p}_{e}-\bm{p}_{\gamma})^2/c^2,
\end{aligned}
\label{def:MM2}
\end{eqnarray}
where $\bm{h}$ is the $\pi^+ \pi^- \eta$ or $ K^+ K^- \pi^0$, $E_{e}$ ($\bm{p}_{e}$) is the energy (momentum) of the positron candidate.
To improve the resolution, we correct the momenta of the ST $D^-_s$ and the transition photon based on the 3C kinematic fit.
To suppress the main background from the decay $D_s^+\to\phi e^+\nu_e, \phi\to K^+K^-$, we require the optimized selection $M_{K^+K^-} > 1.03$ GeV/$c^2$ for the decay $D^+_s\to\ f_1(1420)e^+\nu_e$. Moreover, the optimized selection of $M_{K^+K^-\pi^0e^+}> 1.92$ GeV/$c^2$ is required to suppress the background $D_s^+\to K^+K^-\pi^+\pi^0$ due to $\pi$ and $e$ misidentification.

\section{Results}
\label{subsec:result}
\bigskip

We perform unbinned maximum likelihood fits to the two-dimensional
distributions of $M_{\pi^{+}\pi^{-}\eta}$ ($M_{K^{+}K^{-}\pi^{0}}$) and $MM^{2}$ 
  to search for the 
decay $D_s^+\to f_1(1285)e^+\nu_e$ ($D_s^+\to f_1(1420)e^+\nu_e$).
 The signal probability density functions (PDFs) are
derived from signal MC simulation, while the background PDFs are
modeled using inclusive MC samples; both signal and background yields
are free parameters in the fits. The fit to the $f_1(1285)$ channel
yields $-15.0_{-5.3}^{+6.5}$ signal events, 
while the fit to the $f_1(1420)$ channel yields
$13.2_{-5.0}^{+5.8}$ signal events with a statistical significance of
3.4$\sigma$, as shown in Fig.~\ref{2dfit:1420}. The product BF is
determined to be $\mathcal{B}(D^+_s\to\ f_1(1420)e^+\nu_e)\cdot\mathcal{B}(f_1(1420)\to\ K^+K^-\pi^0) = \rm (4.5^{+2.0}_{-1.7}(stat) \pm0.4(syst)) \times 10^{-4} $, with details of the systematic uncertainties discussed in Sec.~\ref{subsec:BF}.

Given the absence of signal in the decay $D^+_s\to\ f_1(1285)e^+\nu_e$ and the statistical significance of the decay $D^+_s\to\ f_1(1420)e^+\nu_e$ is below 5$\sigma$, we set the upper limits on their BFs, attributing the fitted yields entirely to the $f_1(1285)$ or $f_1(1420)$, respectively.

\begin{figure}[htp]
\begin{center}
\includegraphics[width=0.45\textwidth]{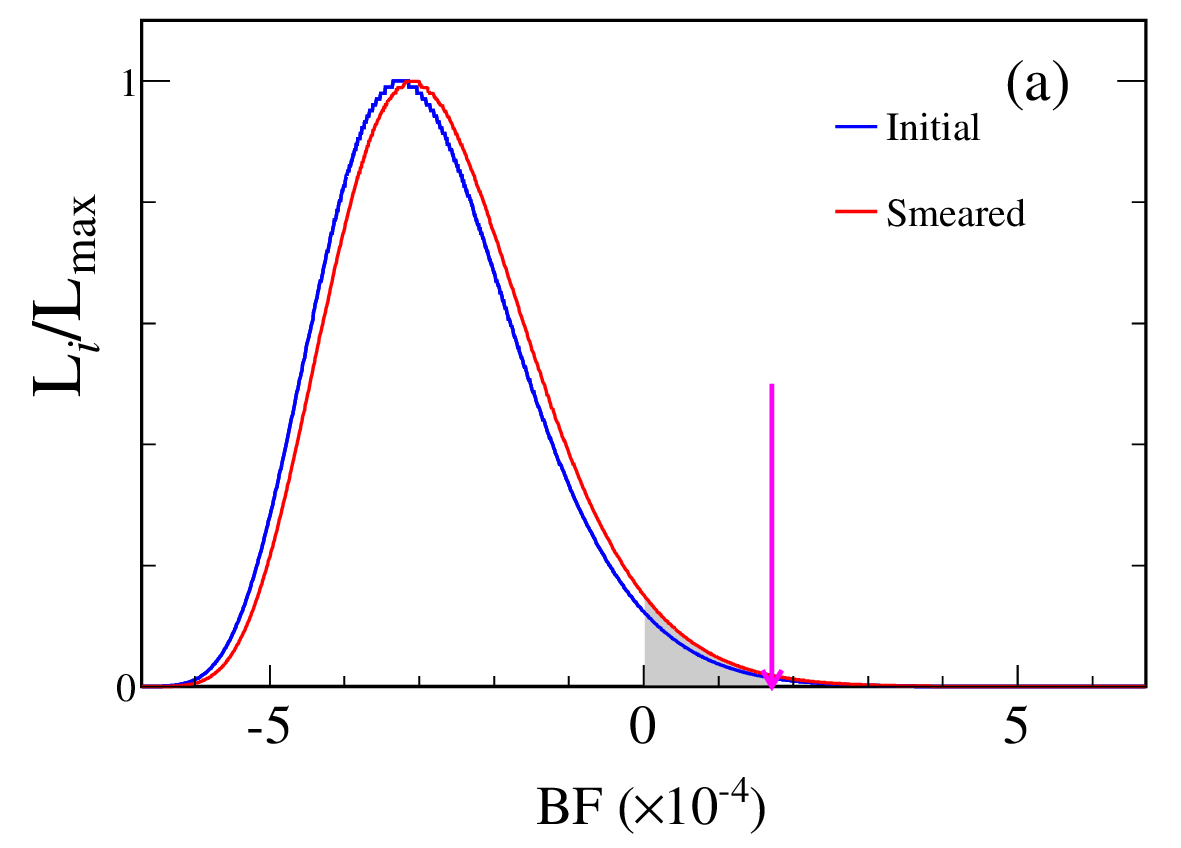}
\includegraphics[width=0.45\textwidth]{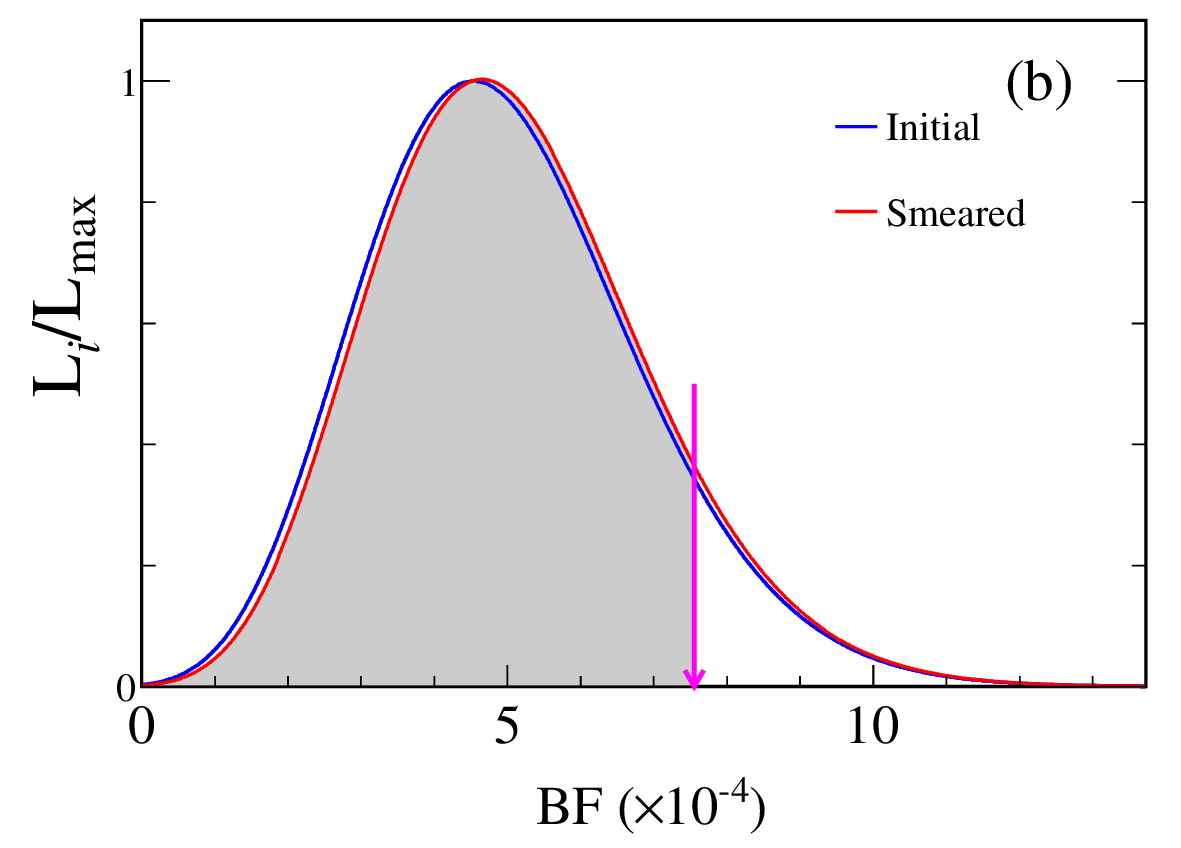}
\caption{
The distribution of likelihood and the corresponding BF products of (a) $D^+_s\to\ f_1(1285)e^+\nu_e, f_1(1285) \to\ \pi^+\pi^-\eta$ and (b) $D^+_s\to\ f_1(1420)e^+\nu_e, f_1(1420)\to\ K^+K^-\pi^0$. The results obtained with and without incorporating the systematic uncertainties are shown in the red and blue curves, respectively. $\rm L_{max}$ is the maximum likelihood  as a function of BF while L$_{i}$ is the arbitrary scan as a function of BF.
The pink arrows correspond to the upper limits at the 90\% confidence level.
}
\label{uplimit11}
\end{center}
\end{figure}

We follow a method described in Ref.~\cite{stenson2006} for incorporating additive and multiplicative systematic uncertainties, which are detailed in Sec.~\ref{subsec:BF}, into the upper limits. First, to take into account the additive systematic uncertainty, the maximum-likelihood fits are repeated using different alternative background shapes and the one giving the most conservative upper limit is chosen.
Second, the likelihood, $L({\cal B})$, as a function of the BF is smeared with the  multiplicative uncertainty $\sigma_\epsilon$ as:
\begin{equation}
L({\cal B}) \propto  \int^1_0 L({\mathcal{B} \frac{\epsilon}{\epsilon_0}}){\rm exp}[\frac{-(\frac{\epsilon}{\epsilon_0}-1)^2}{2\sigma^2_\epsilon}]d\epsilon,
\label{eq:combfbfb}
\end{equation}
where $L({\cal B})$ is the likelihood distribution as a function of the BF; $\epsilon$ is the expected efficiency and $\epsilon_0$ is the averaged MC-estimated efficiency. The averaged efficiencies of $D^+_s\to\ f_1(1285)e^+\nu_e$ and $D_s^+\to f_1(1420)e^+\nu_e$ are $(5.64\pm0.05)$\% and $(1.29\pm0.02)$\%, respectively. The
likelihood distributions incorporating the systematic uncertainties are shown in Fig.~\ref{uplimit11}.  

The upper limits on the product of BFs at the 90\% confidence level, obtained by integrating $L({\cal B})$ from zero to 90\% in the physical region (${\cal B}\leq 0$), are
$\mathcal{B}(D^+_s\to\ f_1(1285)e^+\nu_e)\cdot\mathcal{B}( f_1(1285) \to\ \pi^+\pi^-\eta) < 1.7 \times10^{-4}$,
$\mathcal{B}(D^+_s\to\ f_1(1420)e^+\nu_e)\cdot\mathcal{B}( f_1(1420)\to\ K^+K^-\pi^0) < 7.6 \times10^{-4}$.

\section{Systematic Uncertainties}
\label{subsec:BF}
\bigskip
In determining the BFs, the systematic
uncertainties are categorized as multiplicative and
additive. Multiplicative uncertainties affect the efficiency
determination; Additive uncertainties affect the signal yield
extraction, arising from sources like the modeling of the background
and signal shapes in the maximum likelihood fits.

\begin{table}[htp]
\begin{center}
\caption{Relative systematic uncertainties (in unit of \%) on the BF measurements. Among them, Total (Mul) is only used in the calculation of the upper limit on the BF while Total (Mul and Add) is used in the calculation of the absolute BF.}
\vspace{0.50cm}
\begin{tabular}{l|c|c}
\hline
\hline
Source & $f_1(1285)e^+\nu_e$ & $f_1(1420)e^+\nu_e$\\
\hline
\hline
ST yield                         & 0.3   & 0.3\\
$e^{+}$ tracking                 & 0.5   & 0.5\\
$e^{+}$ PID                      & 0.5   & 0.5\\
$\pi^{\pm}/K^{\pm}$ tracking    & 1.0   & 1.0\\
$\pi^{\pm}/K^{\pm}$ PID         & 1.0   & 1.0\\
MC statistics                    & 0.6   & 0.6\\
$\gamma$ reconstruction          & 2.0   & 2.0\\
$\eta/\pi^0$ reconstruction      & 1.0   & 1.0\\
Signal decay model	                 & 1.7   & 7.3\\ 
${\cal B}(\eta\to\gamma \gamma)/{\cal B}(\pi^0\to\gamma \gamma)$  & 0.5 & 0.03 \\
${\cal B}(D^*_s\to\gamma D_s)$              & 0.4   & 0.4\\
\hline
Total (Mul)                                 & 3.4   & 7.8\\
\hline
Signal shape                                & --    & 1.6\\
Background shape                            & --    & 4.6\\
\hline
Total (Mul and Add)                         & --    & 9.2 \\
\hline
\end{tabular}
\label{tab:Bf-syst-sum}
\end{center}
\end{table}

\subsection{Multiplicative systematic uncertainties}
The multiplicative systematic uncertainties in the measurement of the BFs are listed in Table~\ref{tab:Bf-syst-sum}.

The uncertainty associated with the ST yield is estimated to be 0.3\% by performing alternative fits to the $M_{\rm tag}$ spectra with different signal shapes, background parameters, and fit ranges. The systematic uncertainty related to $e^+$ tracking or PID is assigned as 0.5\% from studies of a control sample of radiative Bhabha events.
The uncertainties associated with $\pi^{\pm}/K^{\pm}$ tracking and PID are investigated using the control sample of $e^+e^- \to K^+K^-\pi^{+}\pi^{-}$. The systematic uncertainty from the tracking or PID is assigned as 0.5\% per kaon and pion, respectively. They are conservatively summed as 1.0\% for $\pi^{\pm}/K^{\pm}$  tracking and PID.

The uncertainty of MC statistics is given by $\sqrt{\sum_{i}{(f_i \frac{\delta\epsilon_i}{\epsilon_i})^2}}$,
where \(f_{i}\) is the fraction of ST yield for the tag mode \(i\), \(\epsilon_{i}\) is the signal efficiency for the tag mode \(i\), and \(\delta\epsilon_{i}\) is its MC statistical uncertainty. This results in an uncertainty of 0.6\%.

For the systematic uncertainty of the transition photon reconstruction, it is assigned to be 2.0\% by studying the control samples of $D^+_s\to K^+K^-\pi^+$ and $D^+_s\to K^0_SK^+$ with the same 3C kinematic fit method. The uncertainties due to the signal decay models are estimated to be 1.7\% and 7.3\% for $D_s^+ \to f_1(1285)e^+ \nu_e$ and $D_s^+ \to f_1(1420)e^+ \nu_e$, respectively, by varying the relative fractions of different $f_1$ subdecays. The systematic uncertainty from the $\eta$ or $\pi^0$ reconstruction is assigned to be 1.0\% ~\cite{PhysRevLett.122.121801}.
The uncertainties in the BFs are 0.5\% for $\eta\to\gamma \gamma$, 0.03\% for $\pi^0\to\gamma \gamma$, and 0.4\% for $D_s^*\to\gamma D_s$, as quoted from Ref.~\cite{PDG2024}.

\subsection{Additive systematic uncertainties}

Additive systematic uncertainty from the background shape is studied by altering the nominal background model obtained from the inclusive MC sample using two methods. First, we generate alternative background models by varying the relative contributions of backgrounds from \(q\overline{q}\) continuum and non-\(D_{s}^{*+}D_{s}^{-}\) open charm processes by \(\pm 30\%\). Second, we change the nominal background shape using the RooKeysPDF kernel estimation function~\cite{RooKeysPdf} with alternative bandwidth parameters $\rho$ = 1, 1.5, 2 and 2.5. Furthermore, To examine the effect of imperfect simulated signal shape, we replace the smooth parameter of the RooKeysPDF function with $\rho$ = 0.5, 1 and 2.5.

In Table~\ref{tab:Bf-syst-sum}, ``Total (Mul)" represents the total multiplicative systematic uncertainty obtained by adding the individual multiplicative components in quadrature. ``Total (Mul and Add)" for \(f_{1}(1420)\) represents the combined systematic uncertainty obtained by adding the total multiplicative uncertainty and the additive uncertainties in quadrature.

\begin{table}[htp]
\begin{center}
\caption{Comparison between our measurement and the theoretical predictions $(\times10^{-4})$. According to the PDG~\cite{PDG2024}, the BF of $f_1(1285)\to\pi^+\pi^-\eta$ is $(35\pm15)\%$, while no BF measurement is reported for the $f_1(1420)\to K^+K^-\pi^0$ decay. After considering the uncertainty of $\mathcal{B}(f_1(1285)\to\pi^+\pi^-\eta)$ from the PDG as part of the multiplicative systematic uncertainty, we obtain the new upper limit.}
\vspace{0.50cm}
\begin{tabular}{l|c|c|c|c}
\hline
\hline
Decay & Ref.~\cite{Cheng2017pcq} & Ref.~\cite{Qiao_2024} & Ref.~\cite{1110930012025} & Our result \\
\hline
$D_s^+\to f_1(1285)e^+\nu_e$ &  [0.6, 3.6]  &   $8.6\pm7.3$   &  $1.17\pm0.18$ &  $ <5.0$    \\
$D_s^+\to f_1(1420)e^+\nu_e$ &  $2.5\pm0.5$ &   $2.1\pm2.1$   &  $3.9\pm0.4$   &  --             \\
\hline
\hline
\end{tabular}
\label{tab:Bf-compare}
\end{center}
\end{table}

\section{Summary}
\label{subsec:summary}
\bigskip
In summary, we present the first studies of the SL decay $D^+_s\to\ f_1(1420)e^+\nu_e,f_1(1420)\to K^+K^-\pi^0$ and $D^+_s\to\ f_1(1285)e^+\nu_e,f_1(1285)\to\ \pi^+\pi^-\eta$ by analyzing 7.33~${\rm fb^{-1}}$ of $e^+e^-$ collision data at $E_{\rm CM}$ between 4.128 and 4.226~${\rm GeV}$ collected by the BESIII detector. The first evidence for the decay $D^+_s\to\ f_1(1420)e^+\nu_e$ is found with a statistical significance of 3.4$\sigma$, and its product branching fraction $\mathcal{B}(D^+_s\to\ f_1(1420)e^+\nu_e)\cdot\mathcal{B}(f_1(1420)\to\ K^+K^-\pi^0)$ is determined to be $\rm (4.5^{+2.0}_{-1.7}(stat) \pm0.4(syst)) \times 10^{-4} $. The corresponding upper limit is determined to be $7.6 \times 10^{-4}$ at the 90\% confidence level, based on the most conservative background assumption.
No significant signal is observed for \(D_{s}^{+}\to f_{1}(1285)e^{+}\nu_{e}, f_{1}(1285)\to \pi^{+}\pi^{-}\eta\), and the corresponding upper limit is determined to be \(\mathcal{B}(D_{s}^{+}\to f_{1}(1285)e^{+}\nu_{e})\cdot\mathcal{B}( f_{1}(1285)\to \pi^{+}\pi^{-}\eta) < 1.7 \times 10^{-4}\) at the 90\% confidence level.
Due to lacking the measured BF of the $f_1(1420)\to K^+K^-\pi^0$ decay, only the result of the $D_s^+\to f_1(1285)e^+\nu_e$ decay is compared with the theoretical predictions, as shown in Table~\ref{tab:Bf-compare}. The comparison shows that our result is compatible with these predictions.
Finally, we hope that the measured BF and upper limits could provide a valuable information for further discussion on the structure of the axial-vector mesons $f_1(1285)$ and  $f_1(1420)$ and the related dynamic of these SL decays.

\section{acknowledgments}
\label{sec:acknowledgement}
\bigskip

The BESIII Collaboration thanks the staff of BEPCII (https://cstr.cn/31109.02.BEPC) and the IHEP computing center for their strong support. This work is supported in part by National Key R\&D Program of China under Contracts Nos. 2023YFA1606000, 2023YFA1606704; National Natural Science Foundation of China (NSFC) under Contracts Nos. 11635010, 11935015, 11935016, 11935018, 12025502, 12035009, 12035013, 12061131003, 12192260, 12192261, 12192262, 12192263, 12192264, 12192265, 12221005, 12225509, 12235017, 12342502, 12361141819; State Key Laboratory of Nuclear Physics and Technology, Peking university under Grant No. NPT2023KFY07; the Chinese Academy of Sciences (CAS) Large-Scale Scientific Facility Program; the Strategic Priority Research Program of Chinese Academy of Sciences under Contract No. XDA0480600; CAS under Contract No. YSBR-101; 100 Talents Program of CAS; The Institute of Nuclear and Particle Physics (INPAC) and Shanghai Key Laboratory for Particle Physics and Cosmology; ERC under Contract No. 758462; German Research Foundation DFG under Contract No. FOR5327; Istituto Nazionale di Fisica Nucleare, Italy; Knut and Alice Wallenberg Foundation under Contracts Nos. 2021.0174, 2021.0299, 2023.0315; Ministry of Development of Turkey under Contract No. DPT2006K-120470; National Research Foundation of Korea under Contract No. NRF-2022R1A2C1092335; National Science and Technology fund of Mongolia; Polish National Science Centre under Contract No. 2024/53/B/ST2/00975; STFC (United Kingdom); Swedish Research Council under Contract No. 2019.04595; U. S. Department of Energy under Contract No. DE-FG02-05ER41374.

\bibliography{ckwx}

\onecolumngrid
\section*{}
\begingroup
\small
\renewcommand{\author}[1]{\begin{center}#1\end{center}}
\author{
M.~Ablikim$^{1}$\BESIIIorcid{0000-0002-3935-619X},
M.~N.~Achasov$^{4,c}$\BESIIIorcid{0000-0002-9400-8622},
P.~Adlarson$^{81}$\BESIIIorcid{0000-0001-6280-3851},
X.~C.~Ai$^{86}$\BESIIIorcid{0000-0003-3856-2415},
C.~S.~Akondi$^{31A,31B}$\BESIIIorcid{0000-0001-6303-5217},
R.~Aliberti$^{39}$\BESIIIorcid{0000-0003-3500-4012},
A.~Amoroso$^{80A,80C}$\BESIIIorcid{0000-0002-3095-8610},
Q.~An$^{77,64,\dagger}$,
Y.~H.~An$^{86}$\BESIIIorcid{0009-0008-3419-0849},
Y.~Bai$^{62}$\BESIIIorcid{0000-0001-6593-5665}
% ...
\\
\vspace{0.2cm}
(BESIII Collaboration)\\
\vspace{0.2cm} {\it
$^{1}$ Institute of High Energy Physics, Beijing 100049, People's Republic of China\\
% ...
$^{4}$ Budker Institute of Nuclear Physics SB RAS (BINP), Novosibirsk 630090, Russia\\
% ...
$^{31}$ INFN Sezione di Ferrara, (A)INFN Sezione di Ferrara, I-44122, Ferrara, Italy; (B)University of Ferrara, I-44122, Ferrara, Italy\\
% ...
$^{39}$ Johannes Gutenberg University of Mainz, Johann-Joachim-Becher-Weg 45, D-55099 Mainz, Germany\\
% ...
$^{62}$ Southeast University, Nanjing 211100, People's Republic of China\\
% ...
$^{64}$ State Key Laboratory of Particle Detection and Electronics, Beijing 100049, Hefei 230026, People's Republic of China\\
% ...
$^{77}$ University of Science and Technology of China, Hefei 230026, People's Republic of China\\
% ...
$^{80}$ University of Turin and INFN, (A)University of Turin, I-10125, Turin, Italy; (B)University of Eastern Piedmont, I-15121, Alessandria, Italy; (C)INFN, I-10125, Turin, Italy\\
$^{81}$ Uppsala University, Box 516, SE-75120 Uppsala, Sweden\\
% ...
$^{86}$ Zhengzhou University, Zhengzhou 450001, People's Republic of China\\

\vspace{0.2cm}
$^{\dagger}$ Deceased\\
% ...
$^{c}$ Also at the Novosibirsk State University, Novosibirsk, 630090, Russia\\
% ...

}
}

\endgroup
\twocolumngrid
\maketitle

\end{document}